\documentclass[12pt,preprint]{aastex}

\usepackage{graphicx}
\usepackage{amsmath}
\usepackage{longtable}
\usepackage{color}

\definecolor{Red}{rgb}{1,0,0}
\definecolor{Green}{rgb}{0,1,0}
\definecolor{forestgreen}{rgb}{0.13, 0.55, 0.13}
\definecolor{Blue}{rgb}{0,0,1}

\newcommand{\be}{\begin{equation}}
\newcommand{\ee}{\end{equation}}
\newcommand{\bea}{\begin{eqnarray}}
\newcommand{\eea}{\end{eqnarray}}

\newcommand{\etal}{et al.}

\newcommand{\btheta}{\mbox{\boldmath$\theta$}}

\begin{document}

\bibliographystyle{apsrev}

\title{A Simple Gravitational Lens Model For Cosmic Voids}

\author{Bin Chen\altaffilmark{1,2}, Ronald Kantowski\altaffilmark{1}, Xinyu Dai\altaffilmark{1}}

\altaffiltext{1}{Homer L. Dodge Department of Physics and Astronomy,
University of Oklahoma, Norman, OK 73019, USA}

\altaffiltext{2}{Research Computing Center, Department of Scientific Computing,
Florida State University, Tallahassee, FL 32306, USA, bchen3@fsu.edu}

\begin{abstract}
We present a simple gravitational lens model to illustrate the ease of using the embedded lensing theory when studying cosmic voids.
It confirms the previously used repulsive lensing models for deep voids.
We start by estimating magnitude fluctuations and weak lensing shears of background sources lensed by large voids.
We find that sources behind large ($\sim$$90\,\rm Mpc$) and deep voids (density contrast about $-0.9$) can be magnified or demagnified with magnitude fluctuations of up to $\sim$$0.05\,\rm mag$ and that the weak-lensing shear can be up to the $\sim$$10^{-2}$ level in the outer regions of large voids. 
Smaller or shallower voids produce proportionally smaller effects.
We investigate the ``wiggling" of the primary cosmic microwave background (CMB) temperature anisotropies caused by intervening cosmic voids.
The void-wiggling of primary CMB temperature gradients is of the opposite sign to that caused by galaxy clusters.
Only extremely large and deep voids can produce wiggling amplitudes similar to galaxy clusters, $\sim$$\rm 15\,\mu K$ by a large void of radius $\sim$$4^\circ$  and central density contrast $-0.9$ at redshift 0.5 assuming a CMB background gradient of $\sim$$\rm10\,\mu K\, arcmin^{-1}$.
The dipole signal is spread over the entire void area, and not concentrated at the lens' center as it is for clusters.
Finally we use our model to simulate CMB sky maps lensed by large cosmic voids. 
Our embedded theory can easily be applied to more complicated void models and used to study gravitational lensing of the CMB, to probe dark-matter profiles, to reduce the lensing-induced systematics in supernova Hubble diagrams, as well as study the integrated Sachs-Wolfe effect.
\end{abstract}

%\pacs{98.62.Sb, 98.80.-k}

\keywords{cosmology: theory---gravitational lensing: weak---cosmic background radiation---large-scale structure of universe}

\maketitle

%--------------------
\section{INTRODUCTION}

The observable Universe is homogeneous but only on very large scales. %($\gtrsim$$200\,\rm Mpc$).
Significant inhomogeneities exist ubiquitously on smaller scales, from galaxies, clusters of galaxies, superclusters, to large cosmic voids.
The large scale structures form the so-called cosmic web, i.e., clusters of galaxies are connected by filaments and walls stretching up to many tens of megaparsecs enveloping vast under-dense voids also having dimensions of tens of megaparsecs (de Lapparent et al.\ 1986; Bond et al.\ 1996; Einasto et al.\ 1997).
Since the discovery of the giant void in Bo$\rm\ddot{o}tes$ (Kirshner et al.\ 1981), cosmic voids are being continually found and investigated (Geller \& Huchra 1989; Peebles 2001; Kamionkowski et al.\ 2009).
Large comprehensive public void catalogues are becoming available thanks to refined void-finding algorithms such as \emph{VoidFinder} (Hoyle \& Vogeley 2004) and \emph{ZOBOV} (Neyrinck 2008) and deep and large redshift surveys such as the Sloan Digital Sky Survey (SDSS, Abazajian et al.\ 2009).
Three recent catalogs are \citet{Sutter12}, \citet{Pan12}, and Nadathur \& Hotchkiss (2014).
These observations show that even though much of the mass in the Universe is bound up in virialized structures such as galaxy clusters, a large fraction ($\gtrsim50\%$) of the volume of the observed Universe is occupied by voids.
Consequently, voids must play an important role in the dynamics of cosmic evolution and in modulating fluxes from a large number of background sources.
Cosmologically distant objects such as supernovae and quasars, and the CMB will be gravitationally lensed by cosmic voids (Amendola et al.\ 1999; Das \& Spergel 2009;  Krause et al.\ 2013; Melchior et al.\ 2014).
For example, the chance for a high redshift SNe Ia (a standard candle) to be lensed by an intervening cosmic void is much higher than being strongly lensed by a foreground galaxy or cluster (Sullivan et al.\ 2010; Clarkson et al.\ 2012; Bolejko et al.\ 2013).
Amendola et al.\ (1999) used the inverted top hat lens model compensated by a thick wall to compute magnification and shear parameters and estimate prospects of detecting weak void lensing.
They concluded by using the color dependent angular density technique (previously proposed to probe cluster densities) that only very large voids could be individually detected.
They also applied the aperture densitometry technique of Kaiser\ (1995) and concluded that a single void lensing signal would be hard to detect because of the large cosmic variance.
Rudinck et al.\ (2007) claimed that to explain the magnitude and angular size of the Cold Spot in the south hemisphere of Wilkinson Microwave Anisotropy Probe (WMAP; Bennett et al.\ 2003) requires a completely empty void ($\delta=-1$) of radius $\sim$$140\,\rm Mpc$ at redshift $z<1,$ far from expectations of the standard cosmology.
Das \& Spergel (2009) constructed a large cosmic void to produce the same Cold Spot.
The model consisted of an uncompensated cylindrical void of comoving radius $150\,\rm Mpc,$ height $200\,\rm Mpc$ and density contrast $\delta=-0.3$ whose axis was along the line of sight.
They concluded that arcminute scale resolution of the CMB would allow detection of their void's  lensing effects.
Krause et al.\ (2013) studied the prospects of constraining the dark matter profile of cosmic voids via the weak lensing effect of stacked voids. By utilizing several compensated lens models, including  the LW12 model of Lavaux \& Wandelt (2012), with $\rho\propto A_0+A_3 r^3$, they concluded that shear and magnification information available from a new generation of spectroscopic galaxy redshift surveys such as Dark Energy Task Force Stage IV surveys, and the Large Synoptic Survey Telescope, would allow sufficient precision to determine the void's dark matter density.
Very recently, Melchior et al.\ (2014) reported the first measurement of reduced gravitational lensing signals arising from cosmic voids by stacking the weak lensing shear estimates around a large number of voids using the SDSS void catalog of Sutter et al.\ (2012). They also used the compensated LW12 model for comparison.
They found that weak lensing suppression is more pronounced near the void radius; however, no useful constraints on the radial dark matter profile were obtained due to the low signal to noise ratio.

Despite the recent observational progress made in searching, characterizing, and classifying cosmic voids, the theoretical modeling of cosmic voids as gravitational lenses has lagged behind.
It has long been conjectured that lensing by under-dense regions in the Universe will give a small dimming effect to background sources: objects behind large empty regions often appear smaller and consequently the observed flux reduced (Sachs 1961; Kantowski 1969; Dyer \& Roeder 1972).
However, no accepted void-lensing model has emerged to date.
One reason is that the traditional gravitational lensing theory was built for mass condensations such as galaxies, instead of under-dense regions such as cosmic voids.
The ambient spacetime of the mass condensation (the lens) is assumed to be flat, a thin lens approximation is assumed, and the lens bending angle is computed by projecting the mass condensation to the lens plane and adding the contribution of each piece linearly.
The above assumptions are adequate for strong, weak, and micro-lensing (Schneider et al.\ 1992), but not for void-lensing, e.g., in the extreme case of a true void, there is no mass to add.
In particular, the traditional lensing theory fails in predicting anti-lensing caused by cosmic voids (Hammer \& Nottale 1986; Moreno \& Portilla 1990; Amendola et al.\ 1999; Bolejko et al.\ 2013).
Another reason is that accurate characterization of void properties, in particular, their sizes and density profiles have only recently become of interest (Sutter et al.\ 2012, 2014; Pan et al.\ 2012; Ceccarelli et al.\ 2013; Tavasoli et al.\ 2013; Hamaus et al.\ 2014; Nadathur \& Hotchkiss 2014).

We have rigorously developed the embedded gravitational lensing theory  for point mass lenses in a series of recent papers (Kantowski et al.\ 2010, 2012, 2013; Chen et al.\ 2010, 2011, 2013) including the embedded lens equation, time delays, lensing magnifications, and shears, etc.
We successfully extended the lowest order embedded point mass lens theory to arbitrary spherically symmetric distributed lenses in \citet{Kantowski13}.
The gravitational correctness of the theory follows from its origin in Einstein's gravity.
The embedded lens theory is based on the Swiss cheese cosmologies (Einstein \& Strauss 1945; Sch$\rm\ddot{u}$cking 1954; Kantowski 1969).
The idea of embedding (or Swiss cheese) is to remove a co-moving sphere of homogeneous dust from the background Friedmann-Lema\^\i tre-Robertson-Walker (FLRW) cosmology and replace it with  the gravity field of a spherical inhomogeneity, maintaining the Einstein equations.
In a Swiss cheese cosmology the total mass of the inhomogeneity (up to a small curvature factor) is the same as that of the removed homogeneous dust sphere.
For a galaxy cluster, embedding requires the over-dense cluster to be surrounded by large under-dense regions often modeled as vacuum.
For a cosmic void, embedding requires the under-dense interior to be ``compensated" by an over-dense bounding ridge, i.e., a compensated void (Sato \& Maeda 1983; Bertschinger 1985; Thompson \& Vishniac 1987; Mart$\rm\acute{\i}$nez-Gonz$\rm\acute{a}$lez et al.\ 1990; Amendola et al.\ 1999; Lavaux \& Wandelt 2012).
%A low density region without such a (compensating) high density boundary has a negative net mass (with respect to the homogeneous background) and is known as an ``uncompensated" void (Fillmore \& Goldreich 1984; Bertschinger 1985; Sheth \& van de Weygaert 2004; Das \& Spergel 2009).
A low density region without a compensating over-dense boundary, or with an over-dense boundary not containing enough mass to compensate the interior mass deficit has a negative net mass (with respect to the homogeneous background) and is known as an ``uncompensated" or ``under-compensated" void (Fillmore \& Goldreich 1984; Bertschinger 1985; Sheth \& van de Weygaert 2004; Das \& Spergel 2009).\footnote{This dichotomy between compensated and un-compensated voids is slightly different from  one based on the classification of the small initial perturbations from which voids are thought to be formed. The initial perturbation can be compensated or uncompensated which leads to different void growth scenarios (Bertschinger 1985), but if the evolved void formed from either perturbation is surrounded by an over-dense shell which ``largely" compensates the under-dense region (i.e., the majority of the void  mass is swept into the boundary shell in the snowplowing fashion when the void is growing), we still call it compensated because the small mass deficit originating in the initial perturbation is unimportant for gravitational lensing.}
Similarly an over-compensated void has positive net mass with respect to the homogeneous FLRW background.
Numerical or theoretical models of over or under-compensated voids do commonly exist (e.g., Sheth \& van de Weygaert 2004; Cai et al.\ 2010, 2014, Ceccarelli et al.\ 2013;  Hamaus et al.\ 2014).
We focus on compensated void models in this paper, given that uncompensated void models do not satisfy Einstein's equations.
The critical difference between an embedded lens and a traditional lens lies in the fact that embedding effectively reduces the gravitational potential's range, i.e., partially shields the lensing potential because the lens mass is made a contributor to the mean mass density of the universe and not simply superimposed upon it.
At lowest order, this implies that the {\it repulsive} bending caused by the {\it removed} homogeneous dust sphere must be accounted for when computing the bending angle cause by the lens mass inhomogeneity and legitimizes  the prior practice of treating negative density perturbations as repulsive and positive perturbations as attractive.
In this paper we investigate the gravitational lensing of cosmic voids using the lowest order embedded lens theory (Kantowski et al.\ 2013).
We introduce the embedded lens theory in Section~\ref{sec:review}, build the simplest possible lens model for a void in Section~\ref{sec:model}, and study the lensing of the CMB by individual cosmic voids in Section~\ref{sec:CMB}.
Steps we outline can be followed for many void models of current interest.

%-------------------------------------------------
\section{FERMAT POTENTIAL AND EMBEDDED LENS THEORY}\label{sec:review}

For spherical lenses we have shown in Chen et al.\ (2013) that the lowest order embedded lens equation can be obtained by minimizing the Fermat potential  (equivalent to the sum of the geometrical and potential time delays, $cT=cT_g+T_p$)
\bea
cT(\theta_S,\theta_I)&=& (1+z_d)\frac{D_dD_s}{D_{ds}}\Bigg[\frac{(\theta_S-\theta_I)^2}{2}
 +\theta_E^2\int_{x}^{1}\frac{f(x',z_d)-f_{\rm RW}(x')}{x'}dx'\Bigg].
\label{T}
\eea
Variation of the image position $\theta_I$ for a fixed source position $\theta_S$ results in the lens equation for an arbitrary mass distribution (Kantowski et al.\ 2013)
\be\label{lens-eq}
y=x-\left(\frac{\theta_E}{\theta_M}\right)^2\frac{f(x)-f_{\rm RW}(x)}{x},
\ee where $\theta_E=\sqrt{2r_{\rm s}D_{ds}/D_dD_s}$ is the standard Einstein ring angle and $\theta_M$ is the angular size of the void, which from embedding is related to the background cosmology by
\be
\theta_M =\frac{1}{1+z_d}\frac{1}{D_d}\left(\frac{r_{\rm s}}{\Omega_{\rm m}}\frac{c^2}{H_0^2}\right)^{1/3}.
\label{thetaM}
\ee
In Eqs.~(\ref{T})--(\ref{thetaM}) $r_{\rm s}$ is the Schwarzschild radius of the FLRW dust sphere's mass removed to form the inhomogeneity; $D_d,D_s,$ and $D_{ds}$ are respectively, angular diameter distances of the lens and source from the observer and the source from the lens; $\Omega_{\rm m}$ and $H_0$ are the matter density parameter and the Hubble constant; $y\equiv \theta_S/\theta_M$ and $x\equiv \theta_I/\theta_M$ are the normalized source and image angles respectively;
$f(x)=M_{\rm disc}(\theta_I)/M_{\rm disc}(\theta_M)$ is the fraction of the embedded lens' mass projected within the impact disc defined by image angle $\theta_I$, and $f_{\rm RW}(x)=1-(1-x^2)^{3/2}$ is the corresponding quantity for the removed co-moving FLRW dust sphere.
Equation (\ref{lens-eq}) is essentially the same as the standard linearized lens equation for spherical mass distributions under the thin lens approximation (Schneider et al.\ 1992) except the $f_{\rm RW}$ term which accounts for the effect of embedding.
At (and beyond) the boundary of the lens, $f(1)=f_{\rm RW}(1)=1,$ the bending angle vanishes and consequently $y=x$,  i.e., the source and image coincide, see Eq.\,(\ref{lens-eq}).
Because the geometrical part of the time delay $T_g$ is universal and only the  potential part $T_p$\, depends on the individual lens structure, all that is needed to construct the Fermat potential is a mass density profile $\rho(x)$ for which
\be\label{Tp}
cT_p(\theta_I,z_d)= 2(1+z_d)r_{\rm s}\int_x^1{\frac{f(x',z_d)-f_{\rm RW}(x')}{x'}{dx'}},
\ee
can be integrated.
The beauty and usefulness of this theory is that all lens properties can be constructed once the specific $T_p(\theta_I,z_d)$  is constructed.
For example the specific lens equation is given by a $\theta_I$-variation $\delta T(\theta_I,z_d)/\delta\theta_I=0$, time delays between image pairs are given by differences in Eq.\,(\ref{T}), and
the integrated Sachs-Wolfe (ISW) effect (Sachs \& Wolfe 1967) is obtained by a $z_d$-derivative  (Chen et al.\ 2013; Kantowski et al.\ 2014)
\be
\frac{\Delta {\cal T}}{{\cal T}}= H_d\,\frac{\partial\, T_p}{\partial\, z_d}.
\label{calT2}
\ee

%--------------------------
\section{AN EMBEDDED VOID LENS MODEL}\label{sec:model}

%Observations (Sutter et al.\ 2012; Pan et al.\ 2012) show that local ($z\lesssim 1$) cosmic voids have very low central densities, $\delta\rho/\bar{\rho}(z_d)\lesssim-0.8$ where $\bar{\rho}(z_d)$ is the cosmic mean at the void's redshift $z_d,$ and are bound by a thin shell of over density.
Observations (Sutter et al.\ 2012; Pan et al.\ 2012) show that local ($z\lesssim 1$) cosmic voids have very low central galaxy counts, $\delta_g\equiv\delta n/\bar{n}(z_d)\lesssim-0.8$ where $\bar{n}(z_d)$ is the mean galaxy number density at the void's redshift $z_d,$ and are bound by an over-dense shell (see e.g., Fig.~9 of Sutter et al.\ 2012).
The thickness of the bounding shell and its density profile are not well constrained by current observations.
The (stacked) void radial profiles are similar for voids of radii from about $10\,h^{-1}\rm\, Mpc$  to about $95\,h^{-1}\rm\, Mpc.$
As a first application of the embedded lens theory to modeling cosmic voids, we construct perhaps the simplest void model, one having a uniform interior density $\rho(x)=(1-\xi)\bar{\rho}(z_d)$  and a thin bounding shell at $x=1$ (of negligible thickness).
The shell contains all the mass removed from the background cosmology to form the under-dense void interior.
Consequently, the void lens model is a strictly compensated one.
Here $0<\xi<1$ is a parameter characterizing the fraction of the background FLRW removed, i.e., the deepness of the void: the larger $\xi$, the deeper the void and the more mass in the shell.
Since the lowest order embedded lens theory depends only on the projected density profile in the lens plane (see Eq.\,(\ref{lens-eq})), the detailed profile of the shell is not important provided that it is sufficiently thin.
Such an inverted top hat model compensated by a thin bounding shell has been proposed by various authors (e.g., Bertschinger 1985; Amendola et al.\ 1999) and has been used to study the ISW effect caused by cosmic voids (Thompson \& Vishniac 1987; Mart$\rm\acute{\i}$nez-Gonz$\rm\acute{a}$lez et al.\ 1990; Chen et al.\ 2013; Kantowski et al.\ 2014).
The infinitesimally thin shell approximation has also been used by other authors to model non-linear void formations/evolutions (Maeda \& Sato 1983a,\,b).

The projected mass fraction function $f(x)$ for this simple void model (uniform interior plus a zero thickness bounding shell) is easily found to be
\be
f(x)-f_{\rm RW}(x)=-\xi x^2\sqrt{1-x^2},
\ee
and the potential part of the time delay, Eq.\,(\ref{Tp}), is
\be
cT_p(\theta_I,z_d)=2(1+z_d)r_{\rm s}\,\left[ -\frac{\xi}{3}\,(1-x^2)^{3/2}\right].
\label{Tp-void}
\ee
By varying $\theta_I$ in the Fermat potential the void's 2D-lens equation is obtained
\bea\label{void-eq}
{\bf y} %&=& x-\left(\frac{\theta_E}{\theta_M}\right)^2\frac{f_{\rm void}(x)+f_{\rm shell}(x)-f_{\rm rw}(x)}{x}\cr
&=& {\bf x}\left[1+\zeta(1-{\bf x}^2)^{1/2}\right],
\eea
where we have introduced a void lens strength parameter
\be\label{zeta}
\zeta\equiv \xi\left(\frac{\theta_E}{\theta_M}\right)^2=\xi(1+z_d)^2\Omega_{\rm m}^{2/3}r_{\rm s}^{1/3}\frac{2D_{ds}D_d}{(c/H_0)^{4/3}D_s}.
\ee
The strength parameter $\zeta$ is proportional to the deepness of the void, $\xi,$ and $r_{\rm s}^{1/3}$ or $\theta_M$ (see Eq.\,(\ref{thetaM})).
The bending angle $\alpha$ (negative/positive corresponds to attractive/repulsive lensing) as a function of the impact $x$ can be read from Eq.\,(\ref{void-eq})
\be\label{alpha}
\alpha(x)
=\xi\frac{2r_{\rm s}}{D_d\theta_M}x(1-x^2)^{1/2},
\ee
where $D_d\theta_I=D_d\theta_M x$ is the physical impact distance.
The Jacobian matrix $A\equiv\partial{\bf y}/\partial{\bf x}$  of the lens equation can be written (Schneider et al.\ 1992)
\be
A=\left(\begin{array}{cc}
1-\kappa-\gamma_1 & -\gamma_2 \\
-\gamma_2 & 1-\kappa+\gamma_1
\end{array}\right),
\ee where the convergence $\kappa\equiv \Sigma/\Sigma_{\rm cr}$ (the projected surface mass density $\Sigma$ normalized by the critical surface mass density $\Sigma_{\rm cr}\equiv c^2D_s/4\pi GD_dD_{ds}$) is
\bea
\kappa &=&-\frac{\zeta}{2}\frac{2-3x^2}{(1-x^2)^{1/2}},
\eea and the two shear components are
\be
\gamma_1= \frac{\zeta}{2}\frac{x_1^2-x_2^2}{(1-x^2)^{1/2}}\>, \hspace{15pt} \gamma_2=\zeta\frac{x_1x_2}{(1-x^2)^{1/2}},
\ee with a total shear
\be
\gamma\equiv\sqrt{\gamma_1^2+\gamma_2^2}=\frac{\zeta}{2}\frac{x^2}{(1-x^2)^{1/2}}.
\ee
The image amplification $\mu$ is
\be
\mu^{-1}\equiv \det |A|=1+\zeta\frac{2-3x^2}{(1-x^2)^{1/2}}+\zeta^2(1-2x^2).
\ee
The tangential component of the shear becomes
\be
\gamma_t\equiv\frac{\Sigma(<\theta_I)-\Sigma(\theta_I)}{\Sigma_{\rm cr}}=\bar{\kappa}(<x)-\kappa(x)=-\gamma,
\ee
where $\bar{\kappa}(<x)=-\zeta\sqrt{1-x^2}$ is the mean convergence within the impact disc at $x$ (the cross shear component $\gamma_\times=0$ for spherically symmetric lenses; Schneider 2006).
The Jacobian matrix $A$ has two eigenvalues
\be
a_t=1+\zeta\frac{1-2x^2}{(1-x^2)^{1/2}}, \>\>\>a_\times=1+\zeta(1-x^2)^{1/2}.
\ee
Eigenvalue $a_t$ can be greater or less than 1 for $x$ less or greater than $\sqrt{2}/2$ (i.e., $\theta_I=\theta_M/\sqrt{2}$) whereas $a_\times$ is always greater than 1.
Consequently, a void-lensed background galaxy is always compressed radially but can be stretched or compressed tangentially depending on the impact.
The lensed image of a small circular source is an ellipse with major axis along the tangential direction ($a_t<a_\times$ for $0<x<1$) and ellipticity $\epsilon\equiv (a-b)/(a+b)\approx- \gamma_t$ ($a$ and $b$ are the lengths of the major and minor axis of the ellipse, respectively).

\begin{figure*}[]
\begin{center}$
\begin{array}{cc}
\hspace{-10pt}
\includegraphics[width=0.55\textwidth,height=0.36\textheight]{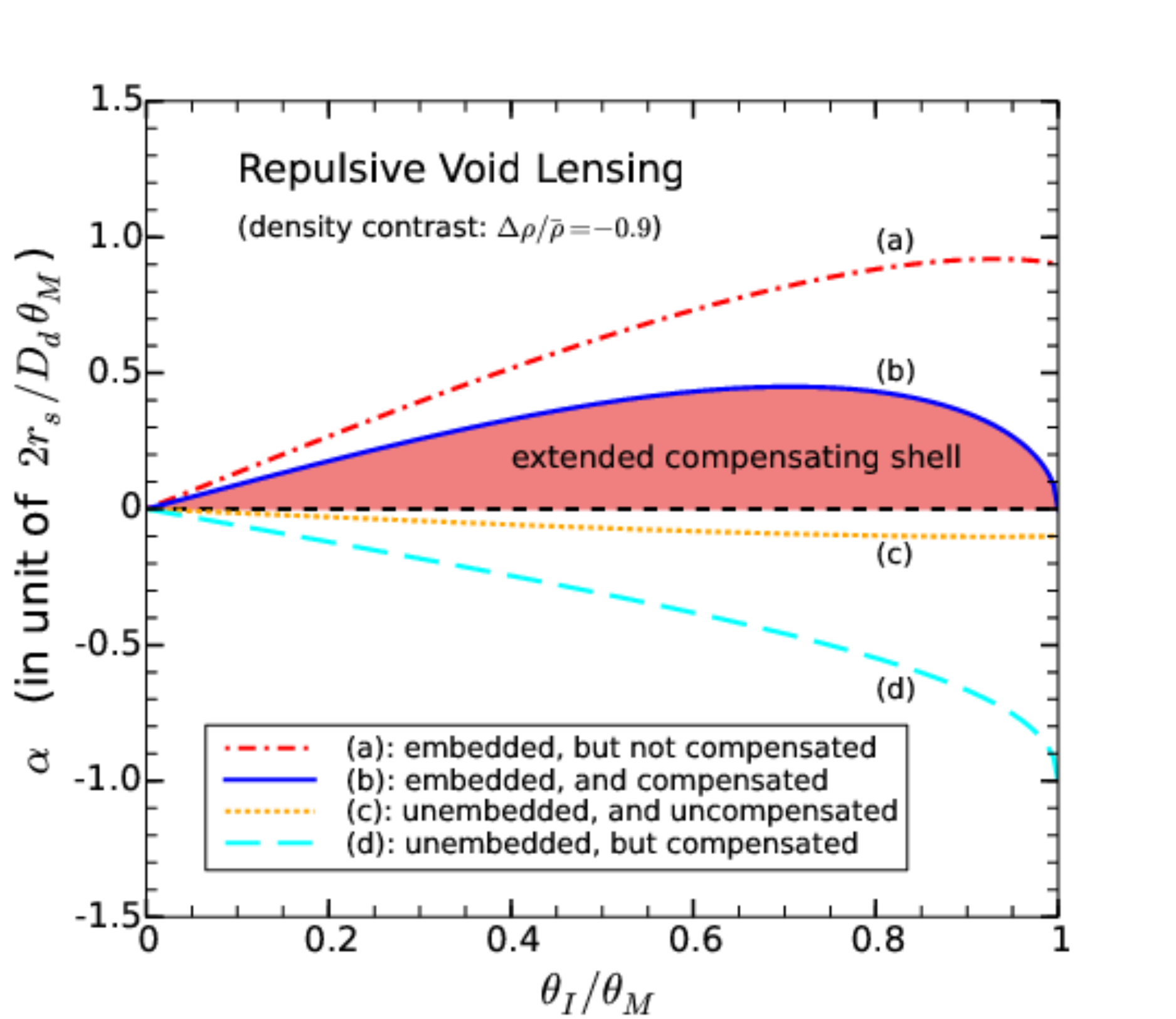}
%\hspace{5pt}
\includegraphics[width=0.55\textwidth,height=0.36\textheight]{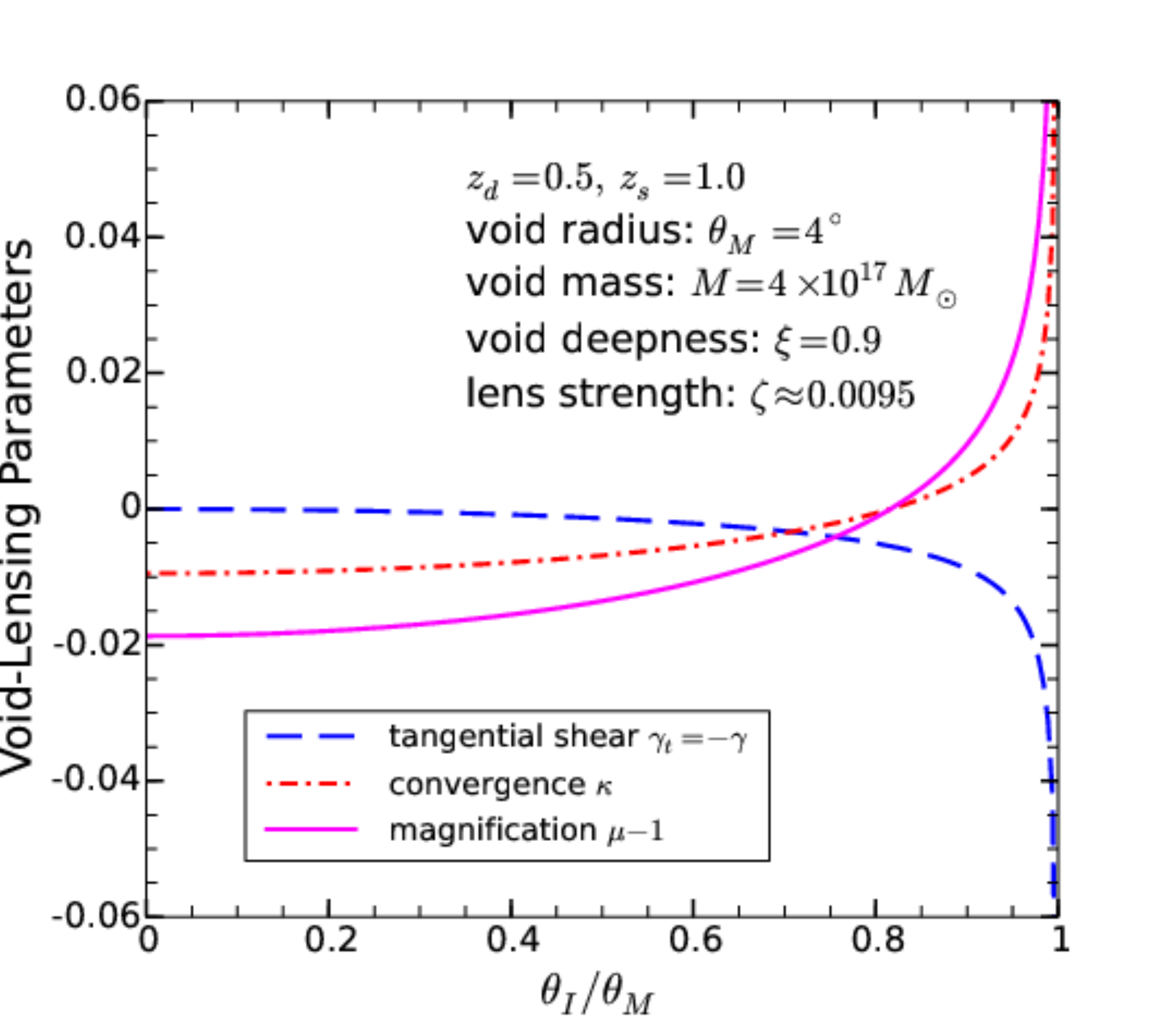}
\end{array}$
\end{center}
\caption{
		(Left) The gravitational bending angle $\alpha$ as a function of the image angle $\theta_I$ for various void lens models.
		For our embedded void lens model (b) we removed a FLRW sphere and replaced it with a uniform void density of $0.1\,\bar{\rho}(z_d)$ bound by a thin shell of negligible thickness at $x=1$ containing $90\%$ of the removed FLRW sphere's mass.
		The bending angle $\alpha$ is the sum of the attractive bending of the under-dense interior and the over-dense thin shell, and the repulsive bending of the removed homogeneous sphere (the effect of embedding).
		Also shown are bending angles for a model (a) which is like (b) but without the presence of the compensating shell;
a conventional lensing model (c) which is like (a) but without the FLRW sphere removed, i.e., a simple 10\% over density; and a conventional lensing model (d) which is like (b) but without the FLRW sphere removed, i.e., a 10\% over-density plus a 90\% shell.
		The bending angle for a lens model possessing an extended compensating shell of finite thickness should be enveloped by model (b) and the x-axis, i.e., lie within the shadowed region.
		Embedded lens theory clearly predicts repulsive bending ($\alpha>0$) for cosmic voids.
		(Right) Void-lensing image magnification $\mu$, convergence $\kappa,$ and shear $\gamma$ as a function of the image angle $\theta_I$ for an embedded and compensated void lens (the blue solid curve in the left panel).
		The observed flux of a source behind a cosmic void can be slightly magnified or demagnified depending on the source position.
		Void-lensing slightly distorts the shape of a background source through a shear $\gamma$ at the few percent level. %(the tangential shear $\gamma_t=-\gamma$ whereas $\gamma_\times=0$ for a spherical void lens).
		The apparent divergence of the lensing parameters near the void boundary is caused by the fact that we assumed a thin shell of negligible thickness for the over-dense bounding ridge.
		The spikes will be shallower and fall back to zero at the void boundary if we assume a thin but finite bounding shell.
			\label{fig:void_lensing}}
     \end{figure*}

%---------------------------------------------------------------------
\subsection{Weak Lensing by A Large and Deep Void}

The simple model we have constructed gives a simple lens Equation (\ref{void-eq}) which contains only one parameter $\zeta.$
The dependence of the lens strength  on the product of the deepness and the radial size of the void is given in Eq.~(\ref{zeta}).
In the following we exhibit properties of such a void assuming it to be of large angular radius $\theta_M=4^\circ$, at redshift $z_d=0.5$, with a deepness parameter $\xi=0.9.$
The physical radius of the void is about $\sim$$88\,\rm Mpc.$
Voids of such angular radii are not unusual, for example, the 50 large voids used in \cite{Granett08} to study the ISW effect have a mean redshift $z_d\sim$0.5 and an angular radius $\sim$4 degrees.
Voids of similar size have also been used to explain the anomalous CMB cold spot on the south hemisphere (Velva et al.\ 2004; Inoue \& Silk 2006; Das \& Spergel 2009; Szapudi et al.\ 2014; Finelli et al.\ 2014; Nadathur et al.\ 2014).
However, voids of radii $\sim$$100\,\rm Mpc$ with dark matter density contrast as large as $-0.9$ are far from being expected in concordance $\Lambda$CDM cosmology.
However, there are structure formation scenarios predicting large and deep voids.
For example, Ostriker \& Cowie (1981) proposed an ``explosive amplification" scenario which can produce large evacuated voids of radius $\sim$$100\,\rm Mpc$ through explosions in the early Universe (see also Bertschinger 1985).
We will use such a large void to study wiggling of CMB primary anisotropies caused by void lensing, and to simulate void lensed CMB sky maps.
As  will be seen, even for an extremely large and deep void, we need very high resolution to resolve  void-bending angles when simulating lensed CMB sky maps.
The  numbers presented in this Section can be used as upper bounds or can be scaled to give results for smaller and/or shallower voids.

Figure~\ref{fig:void_lensing} shows the lensing properties of such a large and deep void.
% of angular radius $\theta_M=4^\circ$ at redshift $z_d=0.5$ with deepness parameter $\xi=0.9.$
The source is at redshift $z_s=1.0$ in a standard flat $\Lambda$CDM background cosmology with $\Omega_{\rm m}=0.3$ and $\Omega_\Lambda=0.7.$
In the left panel we compare the bending angle of the embedded void lens model (b) with those of three other lens models.
Here a void lens model is defined by the density profile of the void and the theory/formalism used to compute the bending angles.
Our simple void model has a uniform low density interior, $\rho(x<1)=0.1\,\bar{\rho}(z_d)$, and an over-dense thin bounding shell containing the remaining $90\%$ of the removed mass, i.e., it is a compensated void.
Our lens model is also an embedded one, i.e., the bending angle $\alpha$ is the sum of the attractive bending of the under-dense interior and the over-dense thin shell, as well as the repulsive bending of the removed homogeneous sphere (the effect of embedding), see model (b) in the left panel of Figure~\ref{fig:void_lensing}.
Model (a) is embedded but not compensated, i.e., the density profile does not include the compensating shell.
Model (c) is neither embedded nor compensated, i.e., the bending angle is simply the attractive bending of the low density interior.
Model (d) is compensated but not embedded, i.e., the bending angle includes contributions from the under-dense interior and the over-dense shell, but not the removed FRLW sphere.
The bending angle is given in unit of $2r_{\rm s}/D_d\theta_M$, about $2.9'$ for the assumed void.
Bending angles of this order of magnitude are comparable to the weak lensing bending angle of the CMB  by large scale cosmic structures ($\sim$$3'$; Cole \& Efstathiou 1989; Seljak 1996b).
Standard lensing theory ignores embedding effect and fails in predicting the repulsive lensing of cosmic voids, i.e., models (c) and (d).
While ignoring the compensating shell  bounding the void, i.e., model (a), significantly over-estimates the magnitude of the bending angle.
There is some evidence suggesting that small voids tend to be over-compensated (e.g., voids in clouds) whereas large voids tend to be under-compensated (Sheth \& van de Weygaert 2004; Cai et al.\ 2010, 2014; Ceccarelli et al.\ 2013; Hamaus et al.\ 2014).
For an under-compensated void lens, i.e., one who's bounding shell contains mass less than $\xi M,$ the bending angle lies between model (a) and (b).
For over or under-compensated lens, the lens potential has an infinite range, whereas for strictly compensated lens, $\alpha=0$ at and beyond the lens boundary.
The embedded lens equation (\ref{lens-eq}) was derived assuming a compensated lens; however, it can be used to model over or under-compensated lens.
But in all cases, the effect of embedding, i.e., the $f_{\rm RW}$ term in Eq.~(\ref{lens-eq}) has to be included to predict  repulsive lensing properties.
A slightly more complex model has a bounding shell of finite thickness whose attractive bending is less significant than the thin shell model assumed in this paper, see Fig.\,3 of Amendola et al.\,(1999).
Consequently, the bending angle of such an embedded and compensated void should be roughly enveloped by model (b) and the x-axis, i.e., the shadowed region.

The right panel of Figure~\ref{fig:void_lensing} shows the void-lensing image magnification $\mu,$ the convergence $\kappa,$ and the tangential shear $\gamma_t=-\gamma$ as a function of the image angle $\theta_I.$
These parameters are determined by the void lens strength parameter $\zeta,$ see Eq.\,(\ref{zeta}). 
For the large and deep voids we have assumed, $\zeta=0.0095.$  
%($\zeta=0.13\%$ for a $10^{15}\,M_\odot$ void of angular radius $0.55^\circ$ at the same redshift,)?????.
Consequently, the convergence, shear, and flux deviation are all at the few percent level, see Eqs. (12), (14) and (15).
A large void of radius $\sim$$4^\circ$ at $z_d\sim0.5$ can introduce a magnitude fluctuation of $\Delta m\approx(-0.05,+0.02)\,\rm mag$ to a background source at redshift $z_s\sim1$.
Fluctuations of this order are consistent with estimates of weak-lensing induced systematics to the supernova Hubble diagram (Kantowski et al.\ 1995; Holz 1998; Wang 2000; Clarkson et al.\ 2012; Fleury et al.\ 2013).
The value of the tangential shear $\gamma_t$ is small except in regions near the void boundary, i.e., $x\gtrsim 0.7.$
This is qualitatively consistent with results of other authors, despite the simplicity of our void model (Amendola 1999; Lavaux \& Wandelt 2012; Krause et al.\ 2013; Melchior et al.\ 2014).
The right panel of Fig.~1 compares favorably with Fig.~3 of  Amendola et al.\ (1999) who used a slightly more detailed lens model with a finite (but narrow) compensating shell.
The apparent divergence of the lensing parameters near the void boundary is caused by the fact that we assumed a thin shell of negligible thickness for the over-dense bounding ridge.
The spikes will be shallower and fall back to zero at the void boundary if we assume a thin but finite bounding shell.
Shear will remain negligible towards the void center even with a finite thickness bounding shell because lensing depends only on the projected mass profile.
Similarly the $\sim$2\% image demagnification near the void center would remain about the same if  we assume a bounding shell of finite thickness. 
The magnitude of the shear parameter we reported is for $\theta_I\approx 0.9\,\theta_M.$
If we replace the razor-thin shell by a shell of thickness, say, $0.1\theta_M,$ the number will be of the same order of magnitude.
The strength of the lens as a whole (e.g., the parameter $\zeta$ for the simple model presented) should not change significantly if we assume a finite thickness bounding shell.
Melchior et al.\ (2014)  made the first measurement of weak void-lensing shear by stacking voids from the Sutter et al.\ (2012) catalog and comparing the theoretical shear profile (Lavaux \& Wandelt 2012) with weak lensing measurements based on SDSS DR8 imaging (Aihara et al.\ 2011).
They found a measured signal for the tangential shear that was also most significant near the void boundary (refer to Figure~2 of Melchior et al.\ 2014).

%--------------------------------------------------------------
\section{GRAVITATIONAL LENSING OF THE CMB BY COSMIC VOIDS}\label{sec:CMB}

Gravitational lensing by intervening large scale structures introduces {\it secondary} anisotropies of the CMB through the ISW effect and remaps the {\it primary} anisotropies through light bending, i.e., CMB lensing (Blanchard \& Schneider 1987; Cole \& Efstathiou 1989; Seljak 1996b; Zaldarriaga 2000; Hu 2001; Okamoto \& Hu 2003; Lewis \& Challinor 2006).
The ISW effect is  caused by evolving inhomogeneities encountered by the CMB after decoupling.
Another important source of secondary anisotropies is the Sunyaev-Zeldovich (SZ) effect (Zeldovich \& Sunyaev 1969; Birkinshaw 1999) which is caused by Compton scattering by hot electrons along the line of sight and should be much less important for low density cosmic voids than it is for high density clusters.
The anisotropy caused by {\it nonlinear} growths of individual mass inhomogeneities at low redshifts ($z\lesssim 2$) was investigated in Rees \& Sciama (1968) using Swiss cheese models and is referred to as Rees-Sciama (RS) effect.
The RS effect has been studied extensively using numerical or analytical techniques such as perturbation theory, N-body simulations, and general relativity (Dyer 1976; Thompson \& Vishniac 1987; Mart$\rm\acute{\i}$nez-Gonz$\rm\acute{a}$lez et al.\ 1990; Seljak 1996a; Cooray 2002a,\,b; Inoue \& Silk 2006; Smith et al.\ 2009; Cai et al.\ 2010, 2014; Hern\'andez-Monteagudo 2010; Ili\'c et al.\ 2013; Finelli et al.\ 2014; Szapudi et al. 2014; Nadathur et al.\ 2014).
Since the RS effect (the late time ISW effect) depends on the evolution rate of the inhomogeneity, a proper {\it embedding} of the lens in the background cosmology is needed to accurately evaluate effects caused by individual clusters or voids.
%Most of the previous work on the RS effect requires a general relativistic treatment and is difficult to follow (Dyer 1976; Thompson \& Vishniac 1987; Mart$\rm\acute{\i}$nez-Gonz$\rm\acute{a}$lez et al.\ 1990).
Recently we have studied the RS effect using the embedded lens theory and obtained a simple analytical expression for the {\it secondary} temperature  anisotropies across an embedded lens using the Fermat potential of the inhomogeneity, see Eq.\,(\ref{calT2}).
We now investigate the lens remapping of the primary CMB anisotropies by cosmic voids using this embedded lensing theory.

\begin{figure*}[top]
\begin{center}$
\begin{array}{cc}
\includegraphics[width=0.5\textwidth,height=0.35\textheight]{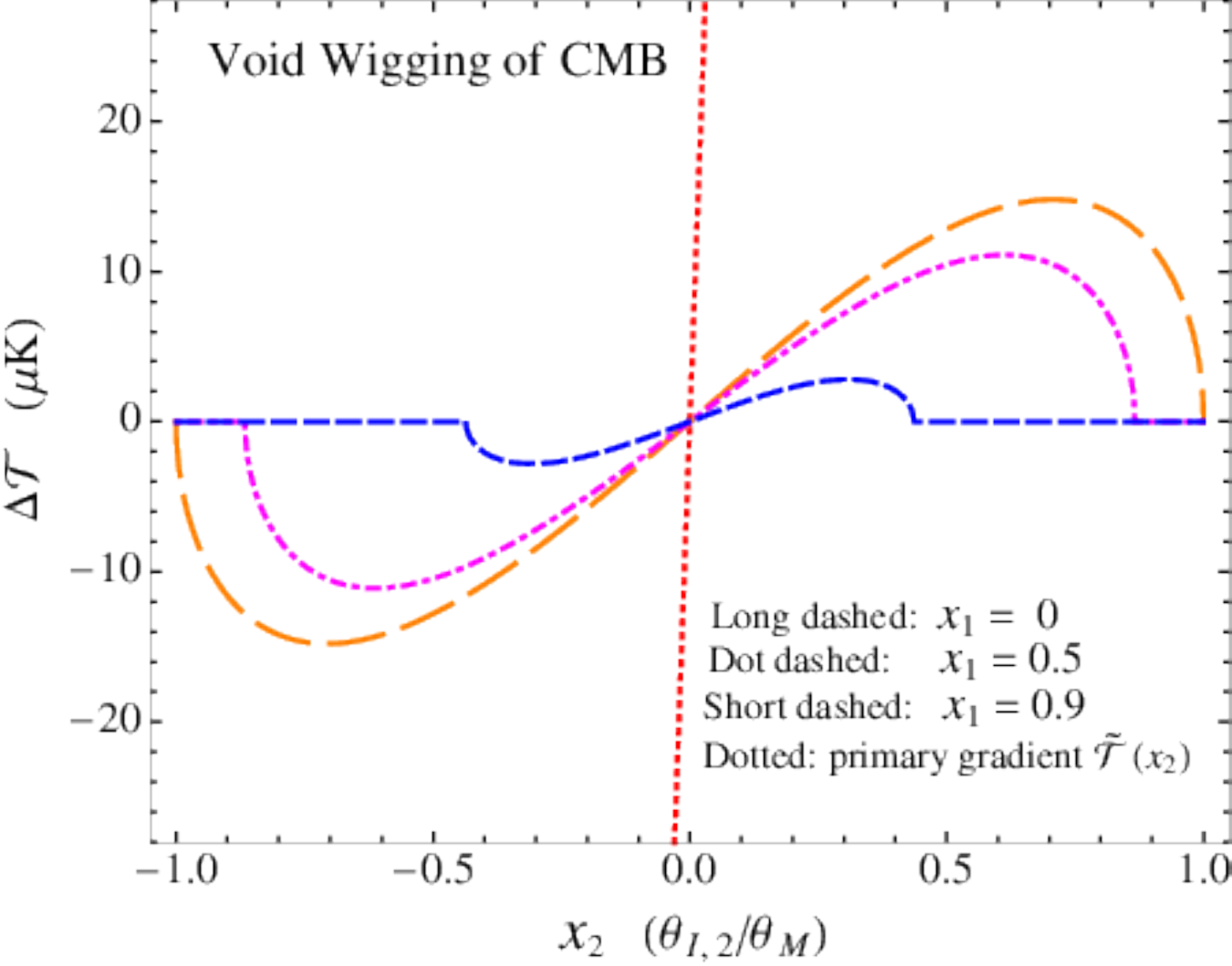}
\hspace{20pt}
\includegraphics[width=0.45\textwidth,height=0.35\textheight]{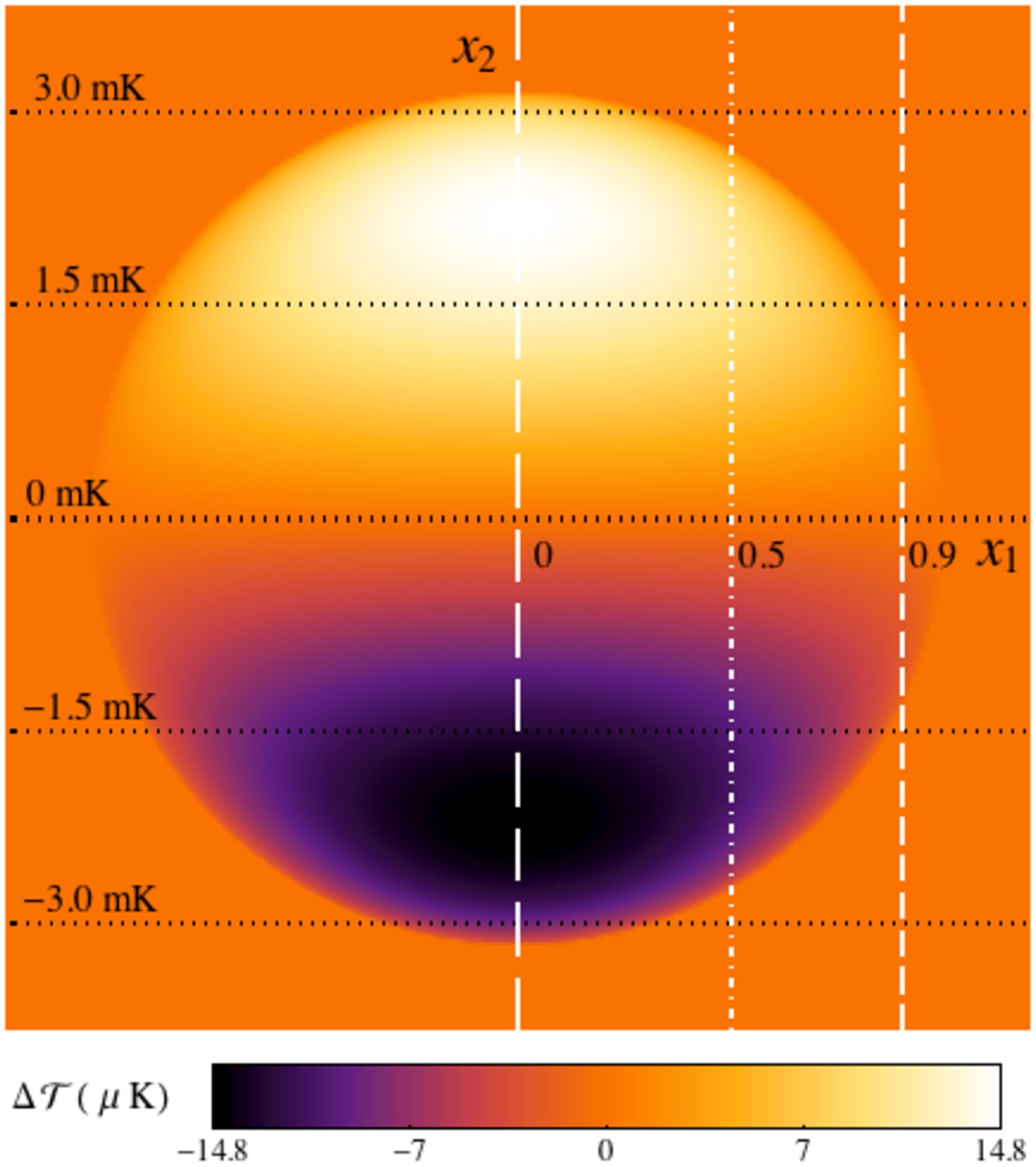}
\end{array}$
\end{center}
\caption{
		 Wiggling of primary CMB temperature gradients caused by a large void lens.
		 The left panel shows $\Delta {\cal T}(\btheta_I)\equiv \Delta {\cal T}(x_1,x_2),$ where $\bf{x}\equiv \btheta_I/\theta_M,$ for fixed $x_1=0,$ $0.5,$ and $0.9$ (the white vertical lines in the right panel).
		 The right panel shows the 2D	temperature perturbation, i.e., $\Delta {\cal T}(x_1,x_2)$.
		 The void model used is the same as in Figure~\ref{fig:void_lensing}.
%		 The primary CMB temperature gradient is taken to be along the vertical axis, $\widetilde{\cal T}(x_1,x_2)={\cal T}_0\,\theta_2$ (the dotted red curve in the left panel, or the dotted black lines in the right panel) with ${\cal T}_0= 13\,\mu \rm K\, arcmin^{-1}$.
		 The primary CMB temperature gradient field is taken to be along the vertical axis, $\widetilde{\cal T}(x_1,x_2)={\cal T}_0\,\theta_2$ (the dotted red curve in the left panel) with ${\cal T}_0= 13\,\mu \rm K\, arcmin^{-1}$ (there is no primary temperature variation along the horizontal direction, see the dotted black lines in the right panel).
		 Void-wiggling of the primary CMB gradient is of the opposite sign and of similar amplitude to that caused by galaxy clusters of similar mass scale (Seljak \& Zaldarriaga 2000).
		The gradient of the void-wiggle, however, is much smaller  than the assumed primary anisotropies, i.e., ($\propto {\cal T}_0/50$).
		 \label{fig:wiggle}}
\end{figure*}

The principle of remapping of primary anisotropies by CMB lensing is simple:  the CMB photons observed at an angle $\btheta_I$ (the image angle) were emitted from a different angle $\btheta_S$ (the source angle) because of lensing.
Consequently, the CMB temperature observed at $\btheta_I$ is different from what it would have been  were there no intervening inhomogeneities.
The reprocessing of primary CMB anisotropies by intervening {\it individual} mass inhomogeneities  has been systematically investigated for mass over-densities such as galaxy clusters (Seljak \& Zaldarriaga 2000; Zaldarriaga 2000; Hu 2001), but not for under-densities, in particular, for cosmic voids.
An exception is \citet{Das09} which studied void-lensing of the CMB assuming a simple uncompensated cylindrical void model of uniform low density whose axis was along the line of sight.
Traditional lensing theory fails in this regime because it fails in predicting the repulsive lensing caused by voids, see Figure~\ref{fig:void_lensing}.
We now estimate the distortions to primordial CMB anisotropies caused by individual cosmic voids using our embedded spherical void lens model.
Assuming the temperature field of the CMB at the last scattering surface had a gradient, we have
\be
{\cal T}(\btheta_I)=\widetilde{\cal T}(\btheta_S)
%=\widetilde{\cal T}(\btheta_I+\delta\btheta)
\approx \widetilde{\cal T}(\btheta_I)+\delta\btheta\cdot\nabla \widetilde{\cal T}(\btheta_I)
\ee
where $\delta\btheta\equiv\btheta_S-\btheta_I$ is the scaled bending angle, and ${\cal T}$ and $\widetilde{\cal T}$ are the lensed and un-lensed temperatures at the respective angles.
For simplicity, we assume the primary CMB temperature field has a large scale gradient along a direction plotted on the $x_2$ axis of the right panel of Fig.\,\ref{fig:wiggle}, with magnitude $\rm13\,\mu K\, arcmin^{-1}.$
Large scale anisotropies of this size are consistent with standard cold dark matter models (Seljak \& Zaldarriaga 2000).
We use the same void model as used in Figure~\ref{fig:void_lensing} and choose $z_s=1100$ as the redshift of the last scattering surface.
The results are shown in Figure~\ref{fig:wiggle}.

In the left panel of Figure~\ref{fig:wiggle}, we plot the void-lensing induced temperature fluctuation, i.e., $\Delta {\cal T}\equiv {\cal T}-\widetilde{\cal T}$ along a few vertical lines across the void (see the white vertical lines in the right panel).
In the right panel, we show the 2-D temperature fluctuation across the void.
The wiggling of the primary CMB temperature gradient can be seen clearly from the left panel of Figure~\ref{fig:wiggle}.
A deep void ($\xi=0.9$) of radius about $4^\circ$ produces  wiggles of amplitude $\sim$$\rm16\,\mu K$ (the amplitude of the wiggling will be down-scaled proportional to the void deepness if the void is shallower). 
The amplitudes of these wiggles are much smaller than the change in the primary CMB temperature field, $\sim$$\rm 3\,m K$ across the void (see the dotted line in the left panel).
A dipole-like feature parallel to the primary temperature gradient is generated by void-lensing, see the right panel of Figure~\ref{fig:wiggle}.
The wigging (the dipole-like feature) is of opposite direction to that caused by galaxy clusters and is due to the repulsive lensing by cosmic voids.
\citet{Seljak00} found steplike wiggles of amplitude $\sim$$\rm10\,\mu K$ for galaxy clusters of velocity dispersion $\sigma_v\sim\rm 1400\, km\, s^{-1}$ assuming the same background gradient field and SIS/NFW cluster profiles (see also Zaldarriaga 2000).
The cluster-wiggling is seen to be confined to the central regions of the cluster ($\sim$$2$ arcmin from the center)  for the examples shown in Fig.\,3 of \citet{Seljak00} and in Fig.\,2 of \citet{Zaldarriaga00}; however, the void-wiggling shown in Fig.\,\ref{fig:wiggle} is spread over most of the void ($\sim$$160$ arcmin).
The cluster produces a dipole-gradient in the central region that is the same order as the assumed primary gradient, $\rm13\,\mu K\, arcmin^{-1}$,
but because the gradient of the void-induced wiggle is spread throughout the void it is smaller by a factor of 1/50 and correspondingly much more difficult to separate from the primary gradient.
This is not  surprising given the nature of weak void lensing.
%The void signal hence has a  much smaller gradient and will be correspondingly difficult to separate from primary gradients.
%In \citet{Chen13} we estimated the ISW effect caused by a cosmic void to be $\sim$$\rm -10\,\mu K$ for a void of similar size using the embedded lensing theory, consistent with recent measurements based on stacking cosmic voids (i.e., the ``aperture photometry" method; Granett et al.\ 2008; Planck Collaboration et al.\ 2014a).
In \citet{Chen13} we estimated the ISW effect caused by cosmic voids using the embedded lensing theory, and found that only extremely large and deep voids can produce signals of about $\sim$$\rm -10\,\mu K$ consistent with recent measurements based on stacking cosmic voids (i.e., the ``aperture photometry" method; Granett et al.\ 2008; Planck Collaboration et al.\ 2014a).
If large cosmic voids are much shallower (e.g., still in the regime of linear evolution) as predicted by concordance cosmology, then the embedded lens theory predicts values much less significant than the observed signals.
It is interesting to note that the amplitudes of void-lensing induced wiggles are at least of the same order as those caused by the ISW effect.

\begin{figure*}[]
\begin{center}
\includegraphics[width=0.9\textwidth,height=0.5\textheight]{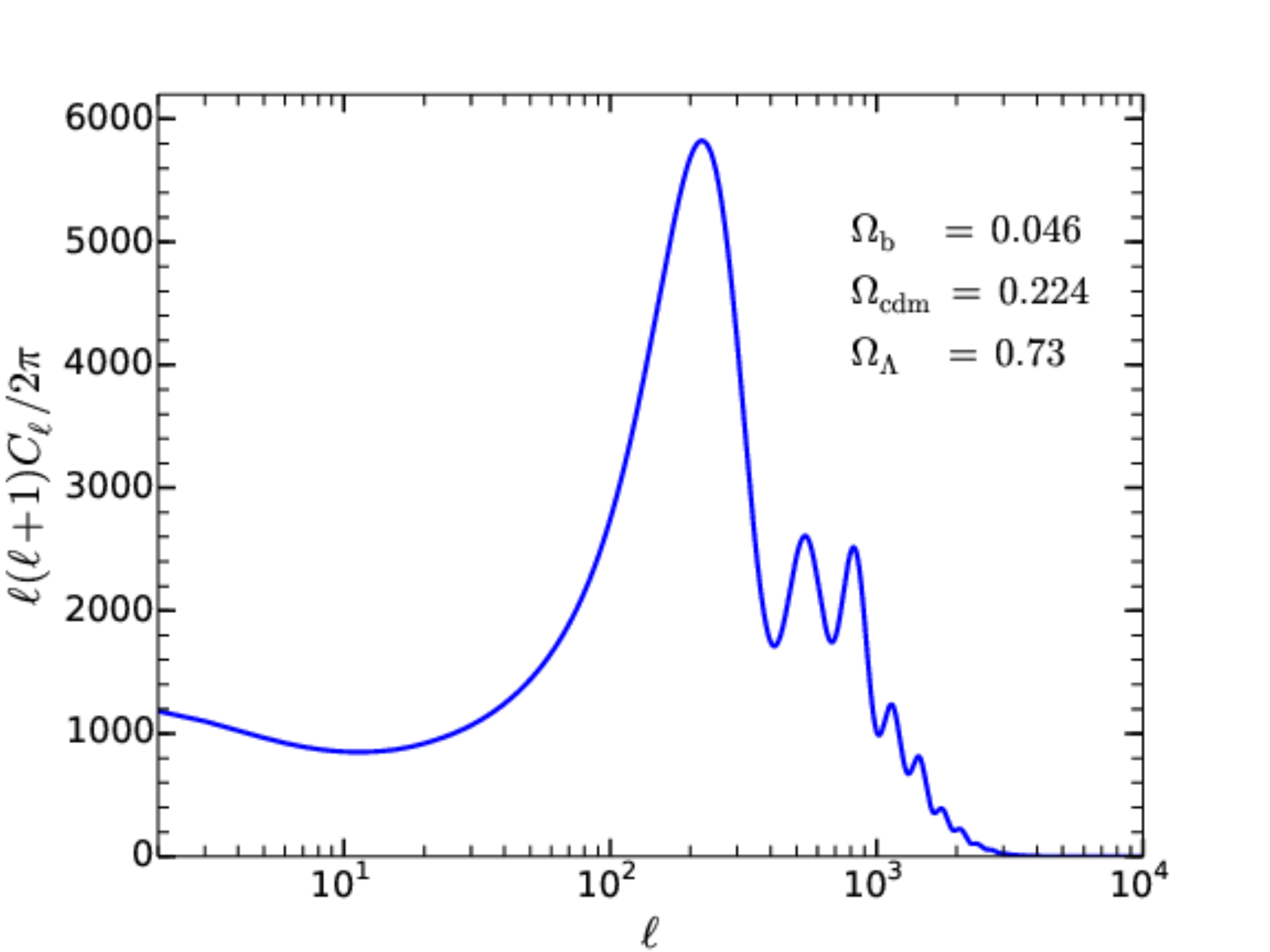}
\end{center}
\caption{ Simulated CMB angular power spectrum using the software package ``CAMB" (Lewis et al.\ 2000).
We have assumed a standard $\Lambda$CDM cosmology with $\Omega_{\rm b}=0.046,$ $\Omega_{\rm cdm}= 0.224$, and $\Omega_\Lambda=0.73.$
\label{fig:power_spectrum}}
\end{figure*}

\begin{figure*}[top]
\begin{center}$
\begin{array}{c}
\includegraphics[width=0.36\textwidth,height=0.29\textheight]{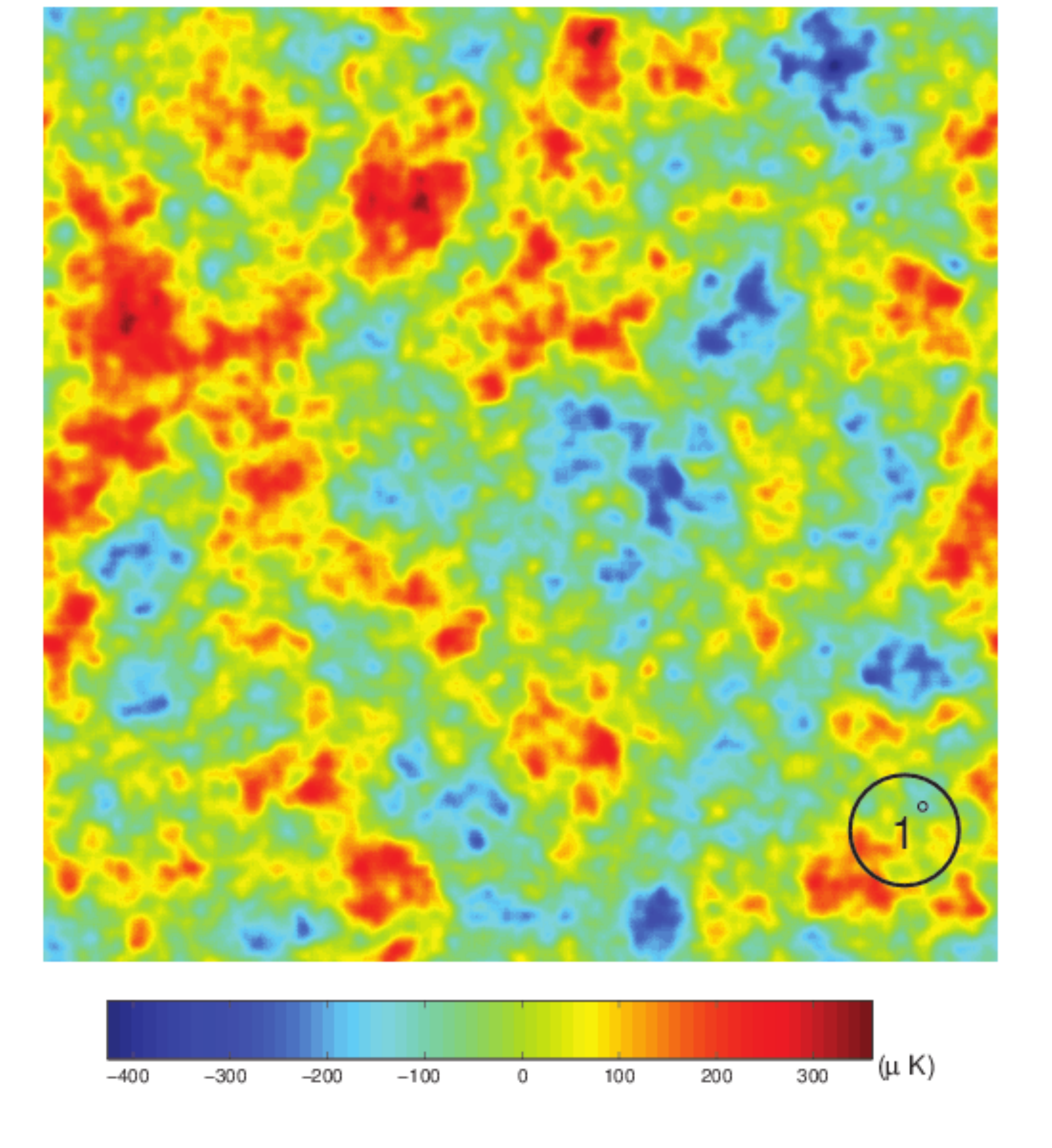}
\hspace{1pt}
\includegraphics[width=0.36\textwidth,height=0.29\textheight]{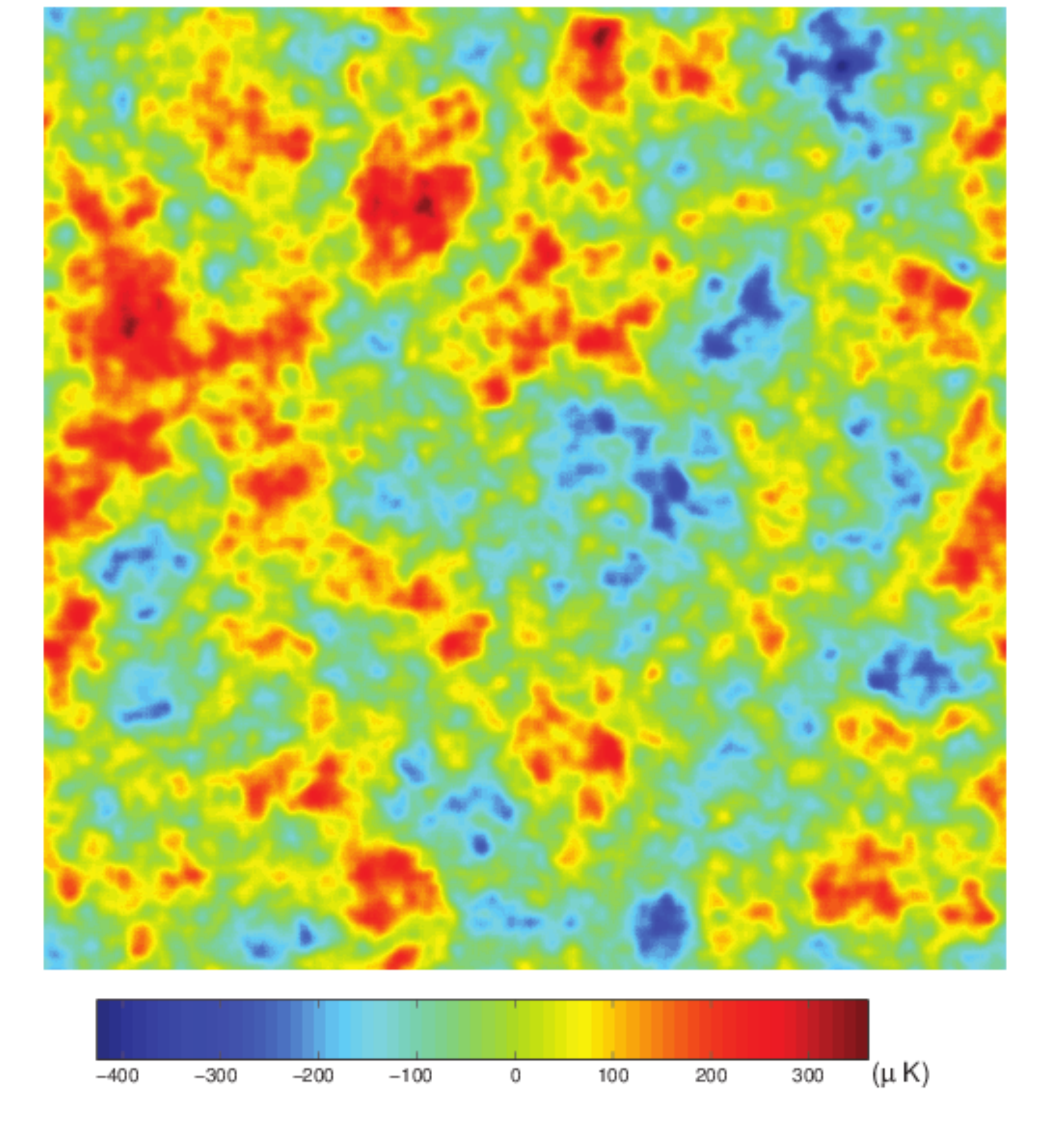}
\hspace{1pt}
\includegraphics[width=0.36\textwidth,height=0.29\textheight]{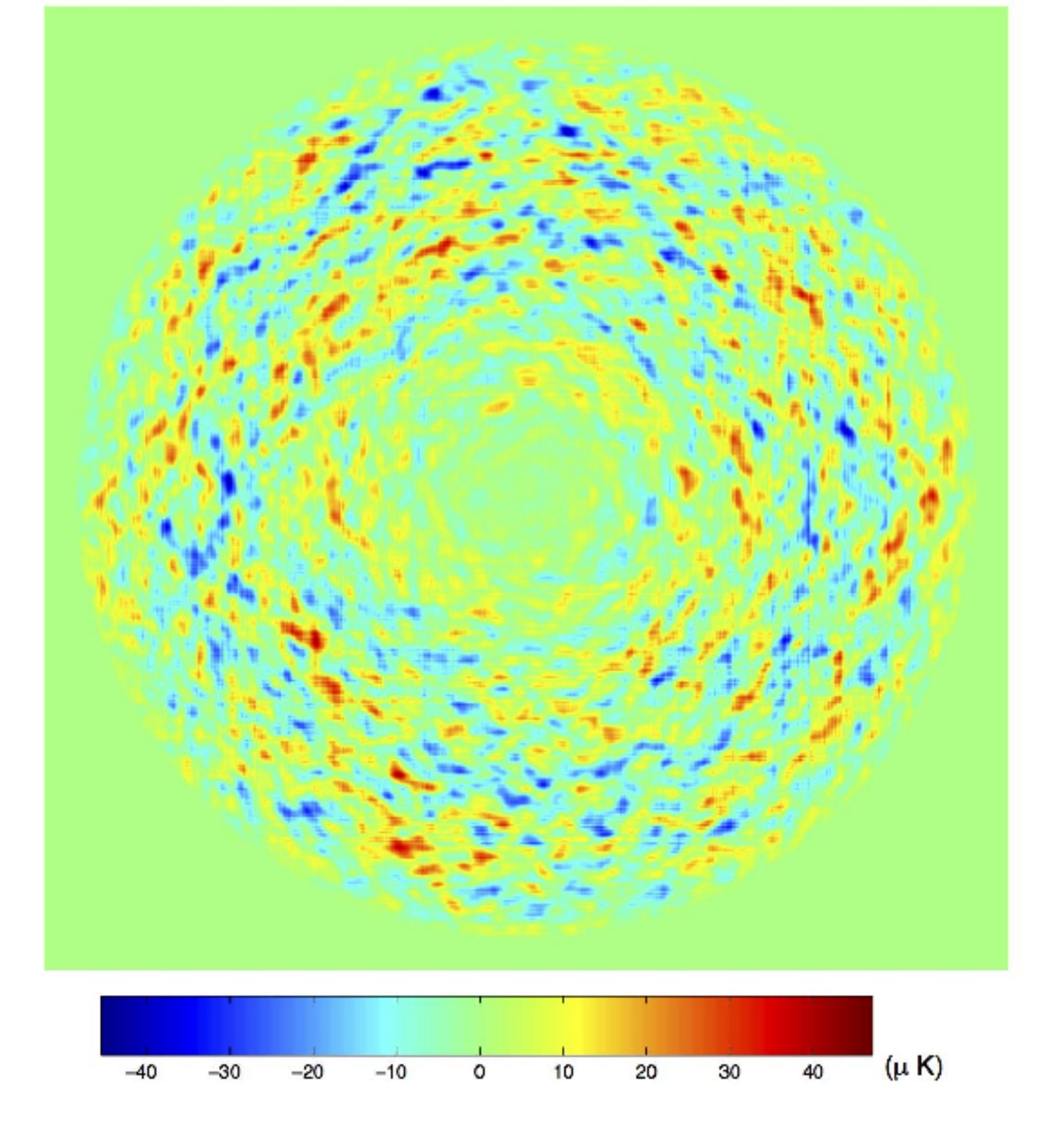} \\

\vspace{15pt}

\includegraphics[width=0.36\textwidth,height=0.29\textheight]{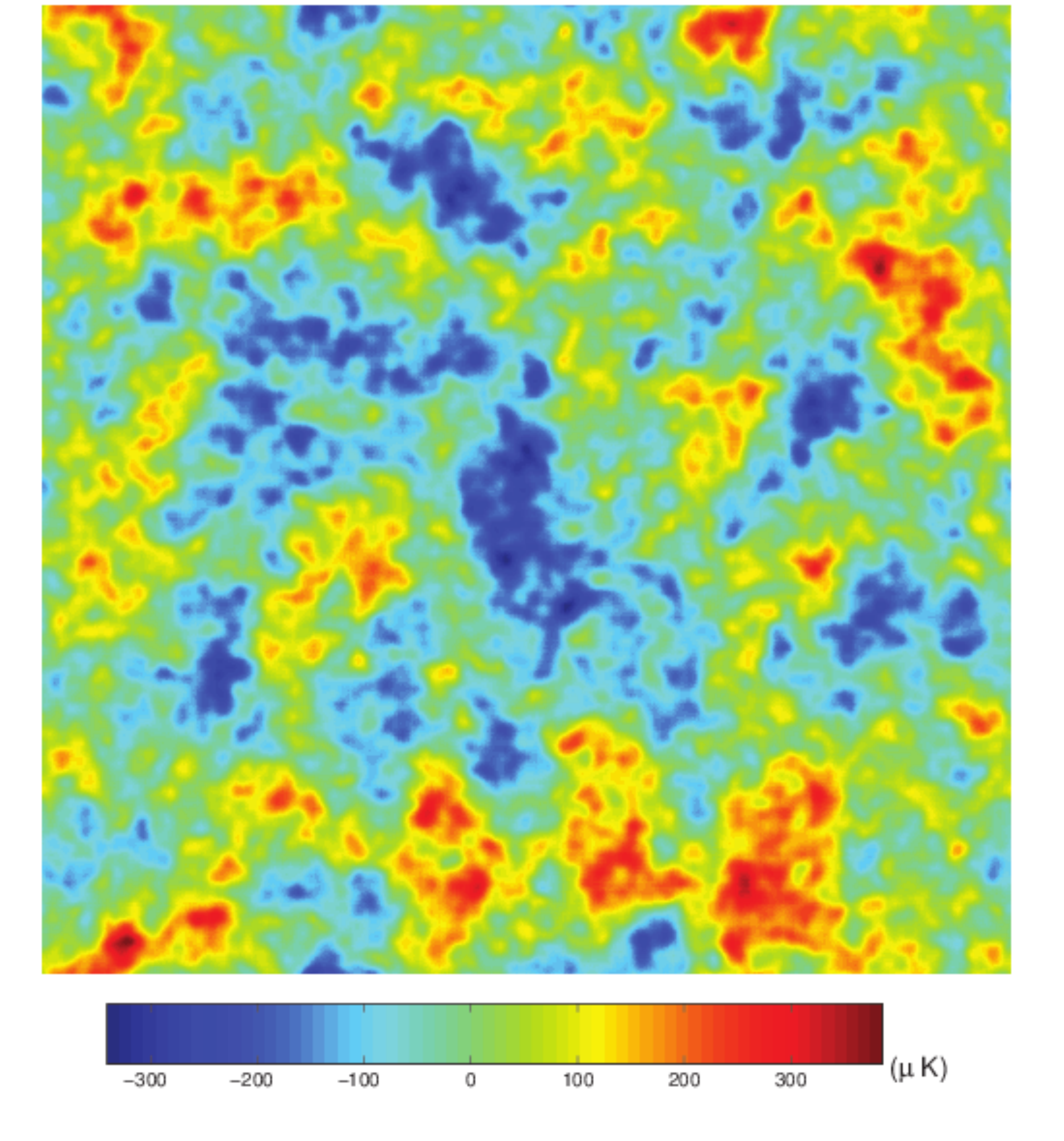}
\hspace{1pt}
\includegraphics[width=0.36\textwidth,height=0.29\textheight]{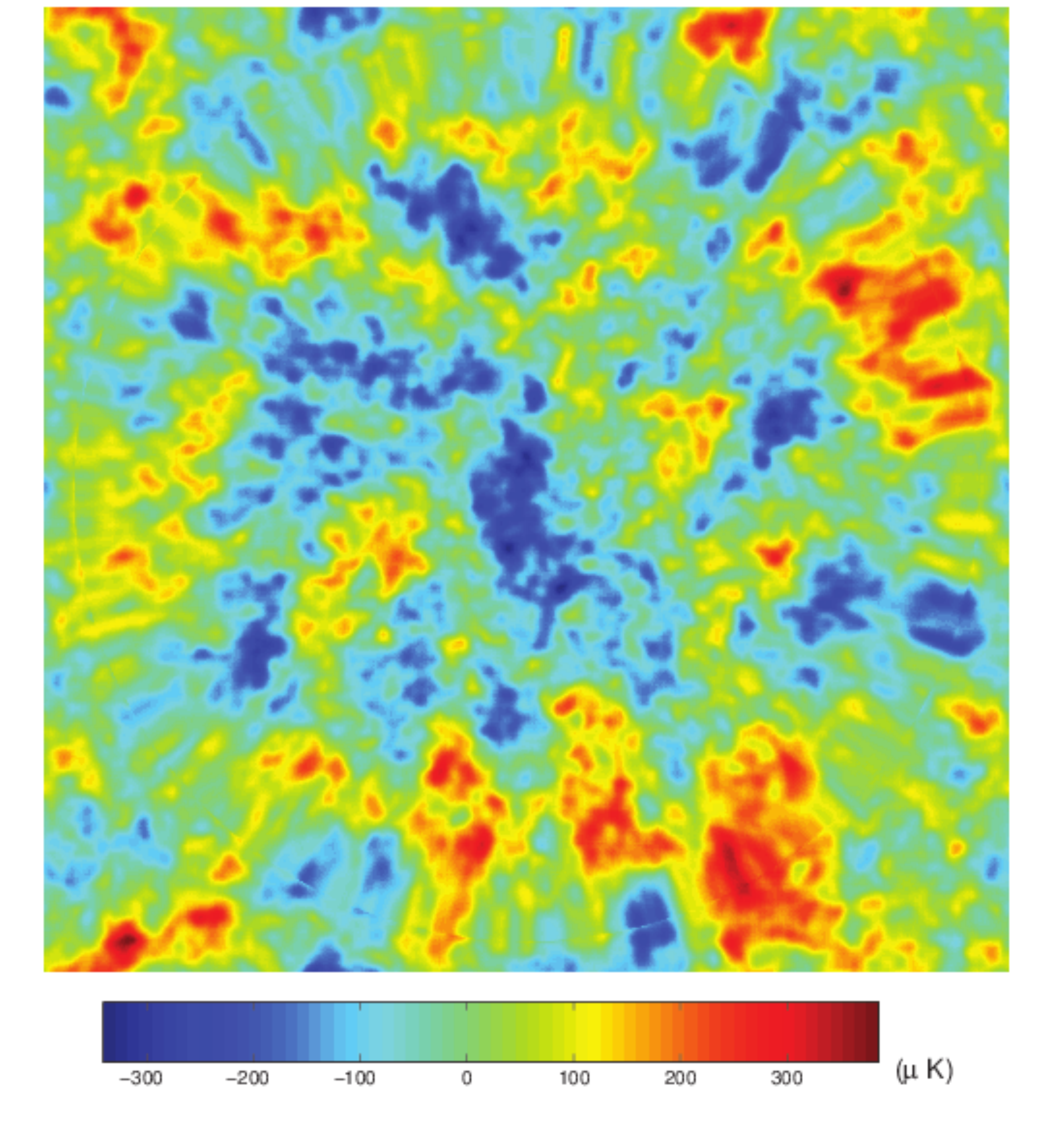}
\hspace{1pt}
\includegraphics[width=0.36\textwidth,height=0.29\textheight]{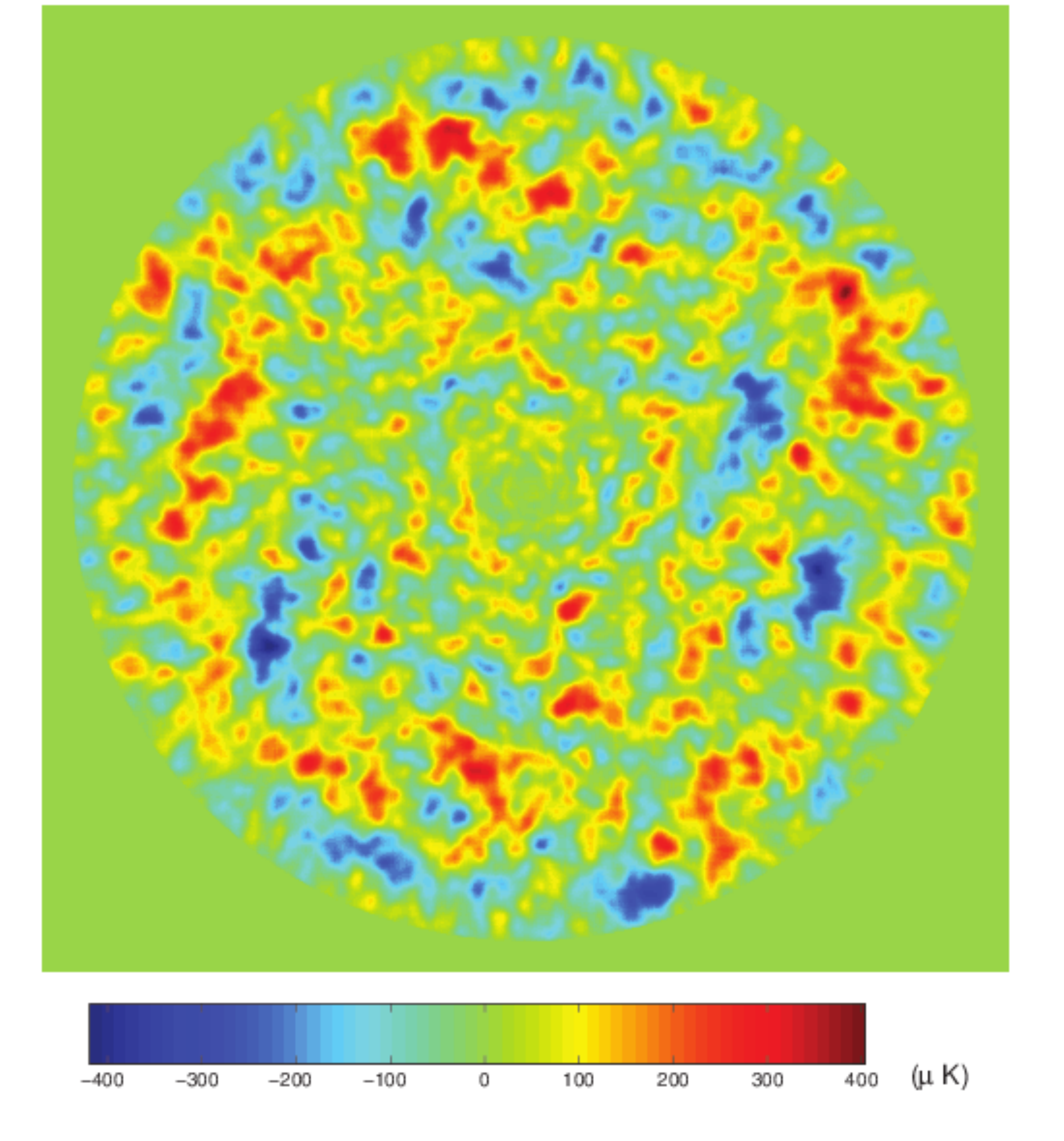}
\end{array}$
\end{center}
\caption{   Simulation of void-lensed CMB maps.
		The first, second, and third columns show respectively the un-lensed CMB map ($8.5^\circ\times 8.5^\circ$), the CMB map lensed by a large and deep void of angular radius $4^\circ$ and central density contrast $-0.9$ at $z_d=0.5,$ and the difference between lensed and un-lensed map.
		The un-lensed CMB map was simulated using the power spectrum shown in Fig.~\ref{fig:power_spectrum}.
		The lens strength parameter is $\zeta=0.0095.$
		For the second row, we amplified the void-lens strength parameter by a factor of 20 ($\zeta=0.19$).
%		For the first row, the GRF has a power-law power spectrum $P(k)\propto k^{-3}.$
%		For the second and third row $P(k)\propto k^{-2}.$
%		For the third row, we amplified the void-lens strength parameter by a factor of 20 ($\zeta=0.19$).
%		The small squares in the center of the difference maps are caused by the fact that bending angle approaches zero near the void center and the discretization errors.
			\label{fig:grf}}
\end{figure*}

The primary CMB anisotropies have small scale structures much more complicated than the simple gradient field assumed in the previous example.
As a second example, we simulate patches of the CMB sky possessing such small structures that are lensed by individual cosmic voids.
We choose the same void model as before with void lens strength parameter $\zeta=0.0095.$
Since the maximum bending angle by a very large and deep cosmic void is only at the arc minute level, the resolution of \emph{Planck}, $\sim$$5'$, is not high enough for our simulation of void-lensing of the CMB.
We simulate high resolution CMB sky maps using the angular power spectrum obtained from the software package ``CAMB" (Lewis \& Challinor 2006) which was based on the CMB code ``CMBFAST" (Seljak \& Zaldarriaga 1996).
We have assumed a standard $\Lambda$CDM cosmology with $H_0= 70\,\rm km\, s^{-1}\, Mpc^{-1} $, baryonic matter $\Omega_{\rm b}=0.046,$ cold dark matter $\Omega_{\rm cdm}=0.224,$ and dark energy parameter $\Omega_{\Lambda}=0.73$.\footnote{Refer to http://lambda.gsfc.nasa.gov/toolbox/tb\_camb\_form.cfm for details.}
Since we want to compute lensing caused by an individual cosmic void, we added `lensing' when computing the CMB angular power spectrum.
To compute the power spectrum without cosmic lensing using CAMB is equally straightforward and will not change any conclusion of this paper.
The simulated CMB angular power spectrum is shown in Fig.~\ref{fig:power_spectrum}.
We first generated simulated  CMB maps of size $8.5^\circ\times 8.5^\circ$ with pixel size $0.125'$ (resolution $4096\times 4096$) using the power spectrum in Fig.~\ref{fig:power_spectrum}.
We then assumed the CMB map was lensed by a void of angular radius $4^\circ$ centering the CMB map and generated the void-lensed CMB map through backward ray-tracing (refer to Eq.\,(\ref{void-eq})).
The results are shown in Figure~\ref{fig:grf}.
The difference maps between the lensed and un-lensed CMB maps are shown in the third column.
As expected, the gravitational lensing distortion to the background CMB map is  weaker than that caused by strong-lensing galaxy clusters. Clusters tend to produce stronger lensing in their central regions, i.e., compare our exaggerated void lens (by a factor of 20) in Fig.\,4 with the exaggerated cluster lens in Fig.\,2 of \cite{Hu01}.
 The void-lensing bending angle is of order  $\sim$$3'$ for the very large and deep void assumed in this paper which has angular radius $4^\circ$ (physical radius $\sim$$88\,\rm Mpc$) and mass $\sim$$4\times 10^{17}\, \rm M_\odot$.
This bending angle occurs at 60--70\% of the voids radius whereas a similar mass cluster has a maximum bending angle limited by the Einstein ring radius (an order of magnitude larger) and occurs near the cluster's center, compare our Fig.\,2 with \cite{Seljak00}.
The majority of the voids found in the local universe have much smaller radii, e.g., of the order $\sim$$10$ Mpc (mass $\sim$$5\times10^{14}\rm M_{\odot}$), and give bending angles   $\sim$$2''$.
If the void is shallower than assumed in this paper ($\xi<0.9$), then the bending angle ($\propto \xi$) will be correspondingly smaller.
On the other hand, the Einstein ring angle of a rich galaxy cluster of $5\times10^{14}\,\rm M_{\odot}$ is of the order $\sim$$1'$.
Correspondingly, the maximum deflection angle of a typical void can be an order of magnitude smaller than that of a condensed rich galaxy cluster.
Given that cluster CMB lensing signals are difficult to detect with current or near-future observations, it will be even more difficult to detect void CMB lensing even if the SZ effect is suppressed \citep{Zeldovich69}.
This by no means implies that the effect of void lensing on the CMB is unimportant.
By statistical analysis of high resolution full/large sky CMB map or by a proper stacking of void-lensed CMB maps, the non-Gaussian signatures of void-lensing might be detected in the future (Zaldarriaga \& Seljak 1999; Okamato \& Hu 2003; Granett et al.\ 2008; Das et al.\ 2011).

%-------------------------------------
\section{DISCUSSION}

We have developed the  simplest spherical void lens model based on the recently developed embedded lens theory.
We have assumed a uniform mass profile for the void, compensated by a thin bounding shell.
The infinitesimally thin bounding shell was chosen for convenience (Maeda \& Sato 1983a,\,b).
To investigate other void profiles such as a non-uniform void interior or a finite-thin bounding ridge (Colberg et al.\ 2005; Lavaux \& Wandelt 2012; Sutter et al.\ 2012; Pan et al.\ 2012; Hamaus et al.\ 2014; Kantowski et al.\ 2014) is straightforward;
one has only to evaluate the Fermat potential of Eq.\,(\ref{T}) or equivalently the potential part of the time delay of Eq.\,(\ref{Tp}).
It is also possible to build embedded void lens models with non-spherically symmetric density profiles given that the lowest order embedded lens theory is applicable to any distributed lens (Kantowski et al.\ 2013).
It is well accepted by the lensing community that small over-densities attract light whereas small under-densities repel light.
This fact can be rigorously proved using general relativistic perturbation theory (Sachs \& Wolfe 1967) assuming $|\delta\rho/\bar{\rho}|\ll 1$.
However, the repulsive nature of lensing by large and deep under-dense region (i.e., cosmic voids) as described by the rigorously derived but simply implemented embedded lens formalism did not appear until Kantowski et al.\ (2013).
In the case of large density contrasts, i.e., $\delta\rho/\bar{\rho}$ approaching its lower bound $-1$ for cosmic voids, the repulsive lens equation follows naturally from the embedded lensing theory. This theory is based on Swiss cheese models (Einstein \& Strauss 1945), which are exact solutions of Einstein's field equations containing inhomogeneities with large density contrasts (Kantowski et al.\ 2010, 2012, 2013; Chen et al. 2010, 2011, 2013).
The void-lensing community takes void repulsive lensing as granted (e.g., Amendola et al.\ 1999; Das \& Spergel 2009), whereas the galaxy/cluster strong lensing community has ignored embedding effects, i.e., the repulsive lensing caused by the large under-dense regions surrounding the central over-dense lens.
Besides correctly predicting repulsive lensing by cosmic voids, our Fermat potential formulation can be used to compute the void-lensing time delay effects including the ISW effect caused by voids, see Eq.\,(\ref{calT2}).

We have used the simplest embedded void model possible to estimate the magnitude fluctuation and the weak lensing shear of background sources lensed by  large cosmic voids, as well as the wiggling of the primary CMB temperature anisotropies.
We estimate that lensing by a deep and large cosmic void can cause flux variations at the few percent level and shear distortions of a background source of a few hundredths.
The possibility for having such a large void of radius $\sim$$100\,\rm Mpc$  and depth $-0.9$ is very small according to concordance $\Lambda$CDM cosmology and hierarchical structure formation theory (Peebles 1980).
The void density profiles extracted by stacking a large number of voids in recent large void catalogs (Sutter et al.\ 2012; Pan et al.\ 2012) suggest very deep void interiors, i.e., $\delta_g\lesssim -0.8$ (refer to Fig.~9 of Sutter et al.\ 2012).
But since these observations are based on galaxy counts, i.e., luminous matter, the actual dark matter density profile can  be much shallower if there exists significant galaxy bias.
However, large and/or deep voids were still proposed as the possible cause of the CMB Cold Spot (e.g., Rudinck et al.\ 2007; Das \& Spergel 2009; Finelli et al.\ 2014; Szapudi et al.\ 2014), and scenarios producing large and deep voids do exist (i.e., the galaxy formation theory via blast waves, see Ostriker \& Cowie 1981).
A galaxy bias $b_g= 1.41\pm 0.07$ was recently used by Szapudi et al.\ (2014) when modeling the CMB Cold Spot using the ISW effect caused by a super giant void.
Similar estimate was made earlier by Rassat et al.\ (2007) for Two Micron All Sky Survey (2MASS) selected galaxies.
This would give estimates of $\delta \equiv\delta_g/b_g\approx -0.6$ assuming a linear bias relation.
The universal void density profile recently proposed by Hamaus et al.\ (2014) using $\Lambda$CDM N-body simulation suggests a deep void interior, $\delta \approx -0.5$ for large voids of radii $\sim$$70\, h^{-1}\rm\, Mpc$ and $\delta \approx  -0.95$ for small voids of radius $\sim$$10\,h^{-1}\,\rm Mpc$ (refer to Fig.~1 of Hamaus et al.\ 2014).
If voids are smaller or large voids are shallower, then our numerical estimates for void lensing image amplification and weak lensing shear should be scaled down accordingly, and the numbers reported here used only as reasonable upper bounds.

Our model predicts a shear distortion that is more significant in the outer region of the voids, qualitatively consistent with recent attempts to detect weak void-lensing shears (Lavaux \& Wandelt 2012; Krause et al.\ 2013; Melchior et al.\ 2014).
We also predict a wiggling of the primary CMB anisotropies at the $\sim$$15\,\mu K$ level assuming a large scale gradient of  magnitude $\sim$$\rm13\,\mu K\, arcmin^{-1}.$
In Fig.~\ref{fig:wiggle} we have simulated for the first time wiggling of the background CMB temperature gradient by cosmic voids.
We also simulated patches of CMB sky lensed by individual cosmic voids and have found that the gravitational lensing distortion to the background CMB map tends to be more distributed throughout the void and less concentrated at its center as for clusters resulting in a much weaker signal in the central region (Seljak \& Zaldarriaga 2000; Zaldarriaga 2000; Hu 2001).
Figure~\ref{fig:grf} contains the first high resolution simulation (pixel size $0.125'$) of the CMB lensed by individual cosmic voids.
If the large void used in this simulation is taken to be much shallower, then a much higher resolution will be  needed to resolve the void's smaller bending because that angle is proportional to the deepness of the void.

We have assumed an infinitesimally thin compensating shell as a crude approximation for simplicity which results in an artificial spike in the shear and amplification at the void's boundary, but since the simulations of the void-wiggling and void lensed CMB sky maps depend only the bending angle, which decreases smoothly to zero for our simple model, the results should be robust. 
Replacing the razor-thin shell by a finite thin one will not change the results significantly (Kantowski et al.\ 2014). 
Embedded void lens models with more physical density profiles should be explored in future work.

To be an exact solution of Einstein's field equations, Swiss cheese models require voids to be strictly compensated, i.e., the added mass in the bounding ridge should exactly cancel the mass deficit in the under-dense interior.
However, theoretical and numerical models of large cosmic voids without compensating shells or with  under-compensated shells are commonly proposed (Sheth \& van de Weygaert 2004; Ceccarelli et al.\ 2013;  Cai et al. 2014; Hamaus et al. 2014).
If such individual voids can be embedded into the homogeneous background, i.e., if their mass density can somehow be a contributor to the background geometry via Einstein's equations without a mass compensation, then predictions made by Swiss cheese embedding may be too conservative as implied by Fig.\,\ref{fig:void_lensing}.

Very recently Melchior et al.\ (2014)  made the first tentative detection of weak void-lensing shear by stacking a large number of voids; however, the signal was rather weak and no useful constraints on the radial dark-matter void profile could be obtained using current data.
Larger void samples and deeper lensing surveys are needed to turn weak void lensing into a powerful probe for the internal structures of cosmic voids (Seljak \& Zaldarriaga 2000; Van Waerbeke et al.\ 2013).
Because of its simplicity of use our embedded void lensing theory can help in ascertaining void density profiles and distinguishing void models.
This theory can also be used to study weak lensing of the CMB by cosmic voids.
The lensing-ISW correlation has been recently detected by \emph{Planck} (Planck Collaboration 2014b).
For example, the CMB void-lensing model presented in this paper can be used in conjunction with the formalism for the ISW effect presented in \cite{Chen13} to study the lensing-ISW correlation bispectrum (Goldberg \& Spergel 1999) caused by cosmic voids.
%Even though such detections are not expected soon,
Such investigations are important given that voids are ubiquitous and the lensing-ISW correlation is the dominate source of secondary non-Gaussianities in the CMB fluctuations and contaminates the measuring of (local) primordial non-Gaussianitie (Zaldarriaga 2000; Planck Collaboration et al.\ 2014b).
Recent constraints on the Hubble constant $H_0$ and matter density parameter $\Omega_{\rm m}$ from \emph{Planck} are at tension with those obtained from SNe Ia Hubble diagram, i.e., the value of $H_0$ and $\Omega_{\rm m}$ from \emph{Planck} are respectively low and high compared with their values inferred from the supernova data (Conley et al.\ 2011; Planck Collaboration et al.\ 2014c).
One possible solution to this apparent discrepancy is the perturbed distance-redshift relation of a clumpy universe (Sullivan et al.\ 2010; Clarkson et al.\ 2012; Fleury et al.\ 2013).
Our theory can be used to reduce the systematics in supernova Hubble diagrams caused by weak lensing of cosmic voids which might help in reconciling the tension between the recent \emph{Planck} results and the supernovae Hubble diagram.

\end{document}